\documentclass[10pt,aps,pra,twocolumn,longbibliography,superscriptaddress,doublespace]{revtex4-2}

\usepackage{graphicx,cancel}
\usepackage{setspace,color,appendix,physics,wasysym,xcolor,xparse}
\usepackage{appendix} 
\usepackage[english]{babel}
\usepackage{amsmath,amssymb,amsfonts,mathtools}
\usepackage{bm,bbm,euscript,braket,esint}
\definecolor{brickred}{rgb}{0.8, 0.25, 0.33}
\definecolor{celestialblue}{rgb}{0.29, 0.59, 0.82}
\definecolor{cornflowerblue}{rgb}{0.39, 0.58, 0.93}
\definecolor{denim}{rgb}{0.08, 0.38, 0.74}
\definecolor{armygreen}{rgb}{0.29, 0.33, 0.13}
\definecolor{cardinal}{rgb}{0.77, 0.12, 0.23}
\definecolor{carnelian}{rgb}{0.7, 0.11, 0.11}
\definecolor{armygreen}{rgb}{0.29, 0.33, 0.13}

\usepackage[colorlinks=true, urlcolor=black, citecolor=blue, linkcolor=red, citebordercolor={1 0 0}, linkbordercolor={0 0 1}]{hyperref}

\usepackage{verbatim,multirow}
\usepackage[utf8]{inputenc}
\usepackage[T1]{fontenc}
\usepackage{times}

\begin{document}

\title{Protocols for counterfactual and twin-field quantum digital signature}
\author{Vinod N. Rao}
\email{vinod.rao@york.ac.uk}
\affiliation{Theoretical Sciences Division, Poornaprajna Institute of Scientific Research, Bengaluru - 562164, India}
\affiliation{School of Physics, Engineering \& Technology, Institute for Safe Autonomy, University of York, YO10 5FT York, U.K.}
\author{Shrikant Utagi}
\affiliation{Department of Physics, Indian Institute of Technology Madras, Chennai - 600036, India}
\author{Anirban Pathak}
\affiliation{Jaypee Institute of Information Technology, A 10, Sector 62, Noida, UP-201309, India}
\author{R. Srikanth}
\email{srik@ppisr.res.in}
\affiliation{Theoretical Sciences Division, Poornaprajna Institute of Scientific Research, Bengaluru - 562164, India}

\begin{abstract}
Quantum digital signature (QDS) is the quantum version of its classical counterpart, and can offer security against attacks of repudiation, signature forging and external eavesdropping, on the basis of quantum mechanical no-go principles. Here we propose a QDS scheme based on quantum counterfactuality, which leverages the concept of interaction-free measurement. Employing the idea behind twin-field cryptography, we show how this two-way protocol can be turned into an equivalent non-counterfactual, one-way protocol, that is both more practical and also theoretically helpful in assessing the experimental feasibility of the first protocol. The proposed QDS protocol can be experimentally implemented with current quantum technology.
\end{abstract}

\maketitle

\section{Introduction \label{sec:intro}}
In contrast to classical cryptography, where the security is due to complexity of the computational problem, the quantum counterpart offers information-theoretic security based on the quantum mechanical principles \cite{bennett2014quantum, gisin2002quantum}. The role of quantum cryptography for varied purposes of communication tasks has been explored extensively in the last four decades. Among them, quantum key distribution (QKD) is the foremost cryptographic task. Several protocols for QKD \cite{bennett2014quantum, ekert1991quantum, bennett1992quantum, guo1999quantum, inoue2003differential, noh2009counterfactual, lo2012measurement, lucamarini2018overcoming} have been proposed and realized in this period (for a review, see Ref. \cite{shenoy2017quantum}). For the present work, a particularly relevant  protocol for QKD (among the above mentioned protocol) is counterfactual QKD proposed by Noh \cite{noh2009counterfactual}.

The concept of interaction-free measurement (IFM), which is the principle behind counterfactuality in certain QKD schemes, involves the counterintuitive idea that quantum superposition can be used to enable the detection of a particle far away from a place where it is blocked \cite{elitzur1993quantum}. The idea of IFM has been exploited for various cryptographic protocols such as key distribution \cite{guo1999quantum, shenoy2013semi, rao2021noiseless}, direct communication \cite{salih2013protocol, aharonov2019modification}, counterfactual universal computation \cite{cao2020counterfactual} and others.

In the prototypical counterfactual QKD protocol (``Noh09'') \cite{noh2009counterfactual}, Alice prepares single-photon states in $\{H,V\}$ basis and sends them sequentially through an unbiased beam-splitter (BS) of a Michelson interferometer. One arm of the interferometer is retained in the Alice's station, whilst the other arm reaches Bob. He may either reflect $H$ polarization while blocking $V$, or vice-versa. Alice and Bob generate a secure key using only those bits when Bob blocks the input polarization and the detections happen at Alice's detector $D_1$ counterfactually. In Noh09, the efficiency is given by $\sum_{j = R_H,R_V} P_{D_1|j}P_j = \frac{RT}{2}$, where $R$ and $T$ are reflectivity and transmittivity with $R+T=1$. When Alice's and Bob's choices are of equal probability, including reflectivity and transmittivity, then one attains the efficiency of 1/8. However, it was recently shown \cite{rao2021noiseless} that by making use of non-counterfactual bits and a simple modification to the original protocol, one may triple the efficiency i.e., up to 3/8. 

The concept of Digital Signature (DS) was first introduced by Diffie and Hellman \cite{diffie1976new}, and could potentially play a crucial role for various cryptographic protocols \cite{gottesman2001quantum, collins2016experimental, nadeem2015quantum, yin2016practical, thornton2019continuous}. A DS protocol involves a sender (Alice) who transmits a digitally signed message $\mathcal{M}$ to the forwarder (Bob), who may forward it to the receiver (Charlie). Even though the message itself is not secret, it should be authenticated, and it needs to be secure against forgery and repudiation. In other words, neither can the sender repudiate her signed message, nor can the forwarder forge or modify the sender's signature if he chooses to forward the message. The main advantage of a DS scheme is that the signed message can be transferred, but cannot be tampered with, so that a third party could also verify the sender's signature and authenticate the message. 

However, since the security of the classical DS schemes is proven by the computational hardness of a mathematical assumption, they cannot offer unconditional security. There lies the advantage of a quantum digital signature (QDS) scheme \cite{gottesman2001quantum}, which utilizes the quantum-secure public keys to validate the message and thus presents an information-theoretic security. Some of the recent works in the area of QDS that exploit various quantum features for security are, the QDS protocol with QKD components \cite{wallden2015quantum}, QDS without the need for quantum memory \cite{dunjko2014quantum}, QDS accounting the difficulties in its practical application \cite{yin2016practical}, QDS without perfect keys \cite{li2023one}, an MDI version of the QDS protocol \cite{puthoor2016measurement}, QDS in a secure network \cite{yin2023experimental}, and others.

The primary security concerns of a (Q)DS protocol are, (a) repudiation by the sender; (b) forgery by the forwarder; and (c) transferability. Repudiation is the act of the sender successfully denying to have sent the message. Forgery refers to the act by an intermediate recipient to forge the signature (i.e., alter the message) of the sender. Another important feature of a QDS protocol is it's transferability, which indicates that if one trusted recipient accepts the message, then another trusted recipient will also accept it if forwarded. Interestingly, in a tripartite scheme, non-repudiation of the message is correlated to its transferability, and can be verified by the mechanism for dispute resolution \cite{amiri2016secure}. In the majority voting for dispute resolution, these two are identical, as a sender who is dishonest will necessarily make the message non-transferable, if repudiation happens. If the sender makes the forwarder accept the message while receiver reject it, it signifies both non-transferability and repudiation. Hence, given a message is rejected by Charlie, it is associated with repudiation by Alice \cite{yin2016practical, thornton2019continuous, roberts2017experimental}. Thus, similar to various other existing QDS schemes, we assume that the receiver is trusted.

The present work is inspired by the idea of utilizing the principle of IFM for a tripartite QDS scheme, and involves effective realization using the setup of the counterfactual QKD protocol. Use of quantum counterfactuality for a QDS scheme provides certain advantages, as the protocol involves only orthogonal states and the experimental setup is simpler \cite{shenoy2017quantum}. Quantum counterfactuality-based QKD protocols have already been experimentally implemented using coherent states \cite{yin2012counterfactual}, thereby making our modified protocol feasible. Finally, the aspect of nonlocality in the context of quantum counterfactuality is interesting \cite{aharonov2020nonlocal}, and our work may potentially lead to studies on tripartite and multipartite scenarios. Note that our work is distinct from various other quantum counterfactual-based three-party protocols, such as certification authorization \cite{shenoy2014counterfactual}, generation of cat states \cite{shenoy2015counterfactual}, quantum key distribution protocol \cite{salih2014tripartite}, and others. In addition, our work also contrasts with other QDS schemes, as we do not require non-orthogonal states \cite{wallden2015quantum, amiri2016secure}, the modified protocol inherently has MDI-like setup \cite{puthoor2016measurement, roberts2017experimental}, the protocol does not require symmetrization \cite{clarke2012experimental}, and it is different from other twin-field based protocols in that we do not have the step of key generation protocol \cite{zhang2021twin}. Furthermore, the requirement of two-way channel is also relaxed in the modified protocol in Sec. \ref{sec:mod}.

The nature of counterfactuality in the direct communication schemes \cite{salih2013protocol, aharonov2019modification, gisin2013optical} is highly debated \cite{salih2014salih, cao2017direct, vaidman2019analysis, hance2021quantum}. Specifically, the issue of weak trace left by the particle is the underlying focus of the debate and thus the need for a more stringent definition of counterfactuality. These schemes are based on the quantum Zeno effect, and they include detections from both the detectors for key generation. However, in the present context, the QDS scheme needs sifting and hence only a subset of detections are used as key bits ($D_1$ detections). Thus the relevance of the argument is limited to only certain counterfactuality-based schemes.

The rest of the paper is structured as follows. In Sec.~\ref{sec:pro}, we present a novel quantum digital signature protocol based on the counterfactual QKD setup. In Sec.~\ref{sec:secu}, we prove the security of the protocol against sender's repudiation and forwarder's forgery. Additionally, we prove the security of the protocol against an eavesdropper's forgery at the level of entanglement in Sec.~\ref{sec:eve}. In practice, the above three-party, two-way, counterfactual QDS scheme is faced with two challenges: the generation of redundant bits lowering efficiency; and, furthermore restricted range due to the requirement of two-way quantum communication. In Sec.~\ref{sec:mod}, we show that this scheme is equivalent to a twin-field setup based one-way QDS protocol which addresses both problems, with certain advantages in deriving secure bounds. Finally, we present our conclusions in Sec.~\ref{sec:conc}.

\section{The protocol\label{sec:pro}}

As noted before, a quantum digital signature protocol would have three stages: distribution, messaging, and forwarding. Let Alice, Bob, and Charlie be the involved parties, who agree on an assigned task: sender Alice transmits the signed message, forwarder Bob verifies the signature and authenticates the message. He may choose to forward the message to Charlie, the receiver, who in turn verifies Alice's signature and authenticates the message. The following assumptions are made in the proposed three-party QDS scheme: (a) The receiver is always trusted; (b) All three parties share an authenticated classical communication channel. Note that these two assumptions suffice for the existence of a QDS protocol with a given number of pre-authenticated parties. However, given that the communication lines are insecure and possibly noisy, the protocol must be made secure against an external Eavesdropper. The authenticated internal parties can detect Eve by measuring the error in the channel. We shall revisit this aspect in Sec. \ref{sec:eve}. The simultaneity in Bob's and Charlie's operations is also assumed to be perfectly timed, along with negligible imperfections in the experimental apparatus. However, since no party colludes with any other, Bob's and Charlie's operations are independent. In some QDS protocols, the message and signature may be made publicly available, but neither can the message be tampered nor can the signature be forged. In our work, we adopt this relaxation and show that the protocol is secure against certain eavesdropping attacks. 

Now, we describe a counterfactual QDS scheme as follows. 

\begin{figure}[ht]
\includegraphics[width=\linewidth]{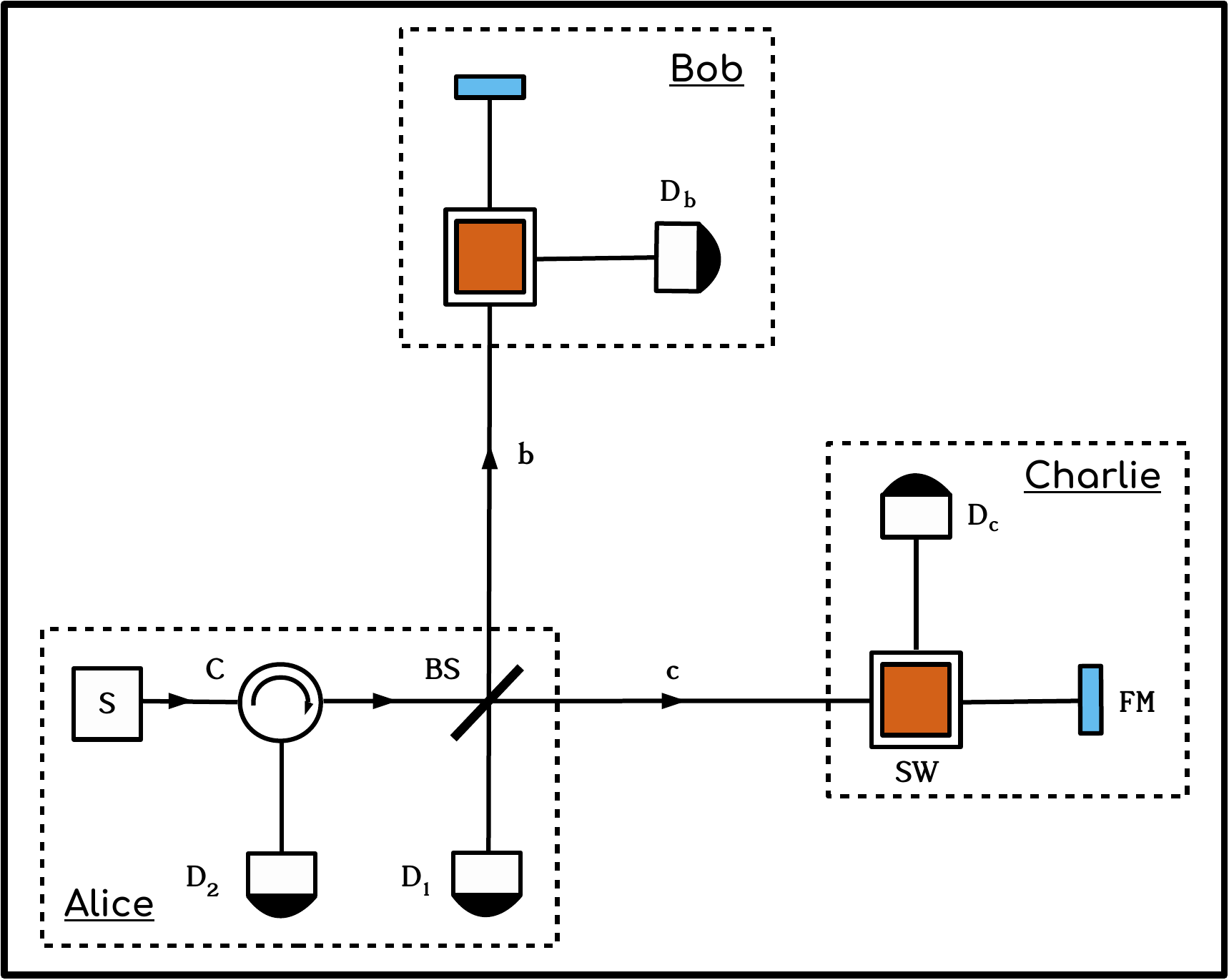}
\caption{Experimental setup for the counterfactual QDS scheme. Alice prepares single-photon orthogonal states and sends it to a beam-splitter (BS) of a Michelson interferometer. At the ends of two arms are Bob and Charlie, who may choose to either block or reflect a polarization. The switch (SW) would have a polarization-BS followed by a circulator (C), for the polarization dependent measurement. The subset of polarization of the photons detected at detector $D_1$ forms Alice's potential signature.}
\label{fig:QDS}
\end{figure}

\subsection*{Distribution stage}
\begin{enumerate}
\item[(D1)] For a future one-bit message $\mathcal{M} = \{m\}$ where ``$m = 0,1$'', Alice prepares a string of $N$ photons in the ${\{H,V\}}$ polarization basis. Each photon is sequentially incident on a beam-splitter (BS) of a Michelson interferometer, with reflectivity and transmissivity being $R$ and $T$, respectively. The end-arms of the interferometer are at Bob's and Charlie's lab, Fig. \ref{fig:QDS}. Let the composite state of $N$ photons be represented by $\ket{\varphi}_m$.

\item[(D2)] Bob and Charlie randomly apply operations $R_H$ or $R_V$, where $R_H$ (resp., $R_V$) represents ``reflect'' $H$ (resp. $V$) and ``block'' $V$ (resp. $H$). A detection could happen at $D_B$ or $D_C$ if their applied operation commutes with the input state. Conditioned on a detection at Bob's/Charlie's station, the respective party may inject a photon of identical polarization towards the BS. This ensures that a given photon detected at $D_1$ is indifferent to the one prepared by Alice. Let $r$ be the fraction of the injection by Bob and Charlie. And independently, on a fraction $f$ for $N$ photons, Bob/Charlie may apply the operation $\sigma_x$ that toggles between the states $H \leftrightarrow V$ for the reflected polarization, where $\sigma_x$ is Pauli-X gate.

The POVMs at the respective detector of Bob's and Charlie's station are $M_H = \text{diag}(0,1,0)$ and $M_V = \text{diag}(0,0,1)$. Similar to Noh09, the respective detection probabilities are, $P(D_B) = R/2, P(D_C) = T/2, P(D_1) = RT/2, P(D_2) = (1+2RT)/4$.

\item[(D3)] The string $\Sigma$ of bits corresponding to Alice's $D_1$ detections forms the sifted key \footnote{Note that in a QDS protocol, the raw key itself can function as the sifted key, since assuming Eve's presence isn't a necessary condition. However, in our protocol, we will indeed prove security under Eavesdropping as well, in which case some bits in $\Sigma$ (check bits) will have to be sacrificed.}. Here, Alice's $D_1$ detections are comprised of (i) counterfactual events; (ii) ones due to non-interference of photon amplitudes (due to both parties reflecting, but only one applying the operation $\sigma_x$); and (iii) Bob's or Charlie's injected photons. Note that (i), (ii) and (iii) are mutually exclusive, and a sifted key bit can be generated if and only if any one of them happens.

\item[(D4)] After Bob and Charlie announce their coordinates of application of $\sigma_x$ to Alice, all the involved parties collaboratively estimate the error in the channel, and if found to exceed an agreed limit, they abort the protocol. Here it is assumed that the BS is unbiased and the fractions $(r,f) \ll 1$.

\item[(D5)] The sifted key $\Sigma$ is of length $(\lfloor\frac{N}{8}\rfloor + \Delta)$, where $\lfloor \cdot \rfloor$ is the floor function and $\Delta$ corresponds to the contributions from non-counterfactual events. Hence no less than $(\lfloor\frac{N}{8}\rfloor)$ of these could be used as Alice's private key. Here, the state $\ket{\varphi}_m$ represents Alice's signature ${\tt Count_{sig}}$, and $\Sigma = \{k_1,k_2,\cdots, k_{N/8}\}$ her private key ${\tt Count_{key}}$. 
\end{enumerate}

\subsection*{Messaging stage}

\begin{enumerate}
\item[(M1)] Alice informs Bob the message $m$, along with her private key ${\tt Count_{key}}$ and the corresponding $D_1$ detection coordinates, in a public channel.
\item[(M2)] For each $k_b^{\rm th}$ bit in Alice's private key ${\tt Count_{key}}$, with $k_b$ denoting bits in $\Sigma$ which Bob knows, he verifies the bit value against his injected bit that led to a $D_1$ detection. He accepts the message if the mismatches are below a threshold.
\end{enumerate}

\subsection*{Forwarding stage}

\begin{enumerate}
\item[(F1)] Should Bob choose to forward the message $m$, we assume that he forwards it to Charlie. If Bob does so, then he also forwards Alice's private key ${\tt Count_{key}}$ to Charlie.
\item[(F2)] Charlie too verifies the Alice's key ${\tt Count_{key}}$ by performing the same procedure described in the Step \textit{(M2)}, but against the set $\{k_c\}$, where $k_c \in \{k\}$ denotes the bits in $\Sigma$ that Charlie knows. This verifies the message, given the mismatches are below the threshold.
\end{enumerate}

Until the messaging stage, the protocol is symmetric with respect to Bob and Charlie. That is, Alice can choose to send $(m,{\tt Count_{key}})$ to either Bob or Charlie and he becomes the forwarder. Below we address the issue of security due to the involved, untrusted parties. The involved parties also estimate error in the channel (in Step (D4)), wherein they may also verify the security of the channel. This is addressed in the Sec. \ref{sec:eve}.

\section{Security against Alice's repudiation and Bob's forgery \label{sec:secu}}
QDS is a cryptographic protocol, wherein the primary security concern is the mistrustful parties and a subset of them could potentially cheat. In the present case of a tripartite scheme, no more than one party is assumed to be dishonest. Message authentication is established by verifying the sender's signature, along with the assumption of an authenticated classical channel shared between parties. Below we address the security of the scheme against Alice's repudiation and Bob's forgery. The transferability of the message can be shown from security against repudiation.

\subsection*{Security against Alice's repudiation}
After Alice sends ($m,{\tt Count_{key}}$), she commits to the message and the private key. Both Bob and Charlie could independently verify her commitment, and we note that the necessity of classical and quantum communication for the same is relaxed here.

Now, Alice's cheat strategies include:
(C.i) Announcing one or more of the $D_1$ detections as $D_2$ detections -- she would necessarily reduce her private key, while not able to change the elements of private key or the signature;
(C.i) Announcing one or more of the $D_2$ detections as $D_1$ detections -- she would be potentially caught as Bob and Charlie can test for such cases against both applying ($R_j,R_j$);
(C.iii) Changing $H \leftrightarrow V$ in $D_1$ -- she would be potentially caught when Bob or Charlie verify against their injected bits.

The optimal cheat strategy for Alice would be to flip the bits from the latter two cases. If successful, this would make Bob accept the given signature bit while making Charlie to reject it. Consider the case in which Bob injects only one bit after a detection at $D_B$, which in turn ends up in $D_1$ detector. There are $\sim 2 \times 2^{(N/8)}$ possible sequences of $\Sigma$ for this single injection case. Similarly, for a higher injected fraction $r$, we notice that the possible sequences for $\Sigma$ is of the order $\sim 2^{(N/4)} \cdot2^{2r}$, accounting for $2r$ injections by Bob and Charlie. 

Suppose she flips a fraction $\tau_A$ in $\Sigma$. Then, from the Chernoff bound \cite{chernoff1952measure}, the success probability of her repudiation is, 
\begin{align}
p(\text{rep}) &= p\bigg[|X - \langle X \rangle_A| \ge \tau_A\langle X \rangle_A\bigg] \nonumber \\
&\le~ 2\exp\left\{\frac{-\tau_A^2\langle X \rangle_A}{3}\right\},
\label{eq:repu}
\end{align}
with $\langle X \rangle_A = N(r_b + r_c)/8$ is the expectation value of the variable $X$ representing the number of $D_1$ detections from injected photons and $r_b$ ($r_c$) is the Bob's (Charlie's) injected fraction. Thus, the probability with which she can escape reduces exponentially, if (a) Alice flips more bits; (b) total number of bits $N$ increases or; (c) the ratio of injected photons increases. Below we shall show why the third case may lead to greater probability of successful forgery by Bob, and hence keep it very low.

\subsection*{Security against Bob's forgery}
Bob's forging action here corresponds to him sending ($m^\prime,{\tt Count_{key}^\prime}$) to Charlie, in a way that Charlie accepts the new message $m^\prime$. Suppose Bob flips a fraction $\tau_B$ in $\Sigma$. Then, similar to Eq. (\ref{eq:repu}), we get the success probability of forgery to be 
\begin{align}
p(\text{forge}) &= p\bigg[|Y - \langle Y \rangle_B| \ge \tau_B\langle Y \rangle_B\bigg] \nonumber \\
&\le~ 2\exp\left\{\frac{-\tau_B^2\langle Y \rangle_B}{3}\right\},
\label{eq:repu}
\end{align}
where, $\langle Y \rangle_B = \Sigma(r_c) - Nr_b/8$, and $\Sigma(r_c)$ signifies $\Sigma$ being a function of $r_c$.

However, in the present context, it can be shown that Bob's best strategy is to flip the bits corresponding to his injected photons in $\Sigma$, because he knows that Charlie is definitely unaware of those injected bits. Then, with unit probability, he succeeds in forging Alice's signature and change the message to $m^\prime$. However, we shall employ classical error correcting scheme to make the protocol robust against Bob's forgery. An alternative solution, but requiring more resource is briefly addressed in Sec. \ref{sec:conc}.

Bob can potentially flip all of his $\big(\frac{Nr_b}{8}\big)$ injected bits during forwarding stage. We include all the injected bits for error correction. We have classical error correction code here of $[n,k,d]$ where $n = \lfloor\frac{N}{8}\rfloor + \Delta$ is the total number of $D_1$ detections, $k$ is the rate of error correction and $d$ is the distance of error in the code. 

We have the following bounds arising from the required error correcting properties. Here we use the notation $n \approx \frac{N}{4}(\frac{1}{2}+r)$, where $r_b=r_c=r$ ($r_c$ is the Charlie's injected fraction) and $n \ge \left|\mathcal{M}\right| = 1$.
The singleton bound \cite{singleton1964maximum} requires $k + d \le n + 1$, which in our case is:
\begin{equation}
N \ge \frac{4\left|\mathcal{M}\right|}{(1-r)} \equiv N_{\rm Sing},
\label{eq:singleton}
\end{equation}
 $\forall$ $k = \left|\mathcal{M}\right|$ and $d \ge \frac{Nr}{2} + 1$.

\begin{figure}[ht]
\includegraphics[width=\linewidth]{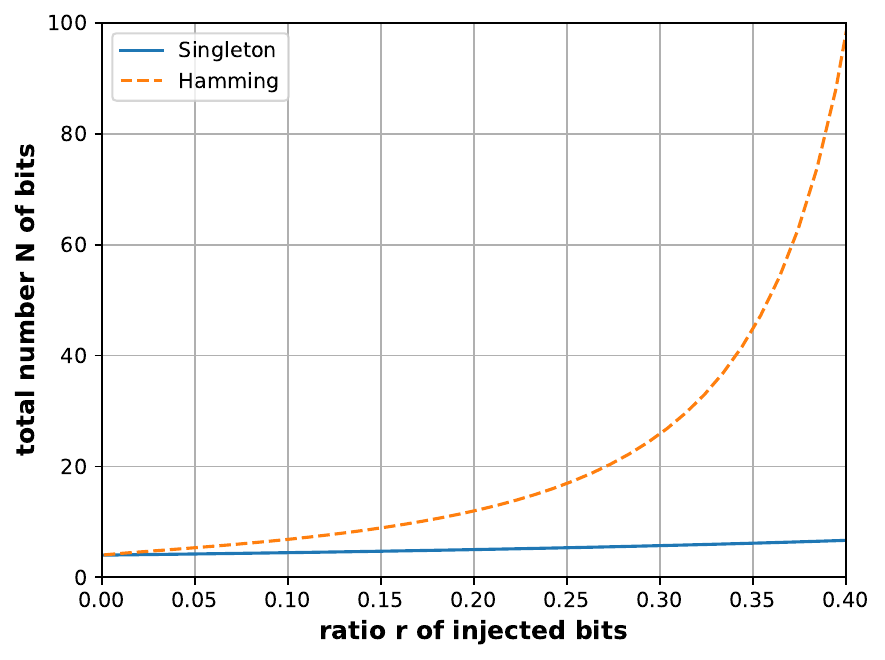}
\caption{Plot of the Hamming and Singleton bounds, Eqs. (\ref{eq:singleton}) and (\ref{eq:hamming}) for the CQDS scheme.}
\label{fig:hamming}
\end{figure}

From Hamming bound \cite{hamming1950error}, $k \le 1 - h(r)$ and thus 
\begin{equation}
N \ge \frac{4\left|\mathcal{M}\right|}{[(1+r)(1-h(r))]} \equiv N_{\rm Hamm},
\label{eq:hamming}
\end{equation}
$\forall$ $k = \frac{\left|\mathcal{M}\right|}{n}$ (rate of error correcting code) and provided $r \le \frac{1}{2}$.
The bounds in Eq. (\ref{eq:singleton}) and Eq. (\ref{eq:hamming}) are plotted in Fig. \ref{fig:hamming}.

We can replace $r^\prime$ with $r$ (where $r^\prime < r$) in both the bounds, as doing so does not change the nature of the value of $N_{\rm min}$. Also as observed in the above graph, the Hamming bound implies a higher $N_{\rm min}/\textrm{bit}$ value at all stages than the Singleton bound, and thus places the actual bound for the protocol. The above passive attack \cite{dunjko2014quantum} involves Bob being honest during distribution stage, and later trying to forge Alice's signature. However, we may consider the active attack, wherein Bob is malicious in the distribution itself. But the underlying security remains same, wherein Alice and Charlie can test for Bob being malicious using the statistics of counterfactual detections. 

\section{Security against eavesdropping \label{sec:eve}}
The discussions so far have assumed that there is no eavesdropper's interference. This is reasonable in a mistrustful protocol, where the players themselves pose the main security threat to each other. However, in a practical situation, an eavesdropper Eve may be present, whose objectives are given later. As noted in Sec. \ref{sec:pro}, the internal members of protocol are oblivious to the presence of Eve, it is natural to assume that no party colludes with Eve. In the case of collusion, there is no specific advantage they get in terms of cheating, assuming that they cheat out of their own self-interest. Therefore, it suffices to prove security against an external, unauthenticated Eve.

Given that the protocol is asymptotically secure against Alice's repudiation and Bob's forgery, in this section, we will show that it is asymptotically secure against eavesdropping. Since we prove security at the entanglement level, it may be presumed that the protocol is secure against more general attacks. Note that the message itself is not the secret, but Alice's signature is. So Eve would try to get information of ${\tt Count_{sig}}$ so that she can tamper with Alice's message later.

The basic yet powerful attack Eve could do is the intercept-resend attack, where she would block/resend a photon during the quantum communication stage. The best possible attack strategy for Eve would be to use the same \textit{reflect} or \textit{absorb} operator at both the arms (i.e., $E_B$ -- Eve's ancilla at arm $b$ and $E_C$ -- Eve's ancilla at arm $c$). Whenever there is a detection at Eve's detector, she would either choose to send the identical photon towards Bob/Charlie or towards Alice. Even when there is no detection, she would get the bit's info, but not induce any error. Thus she can get the complete information of all the bits from this attack with only $50\%$ detections ($50\%$ error). 

As a countermeasure for this, we propose an extra stage. As mentioned after the protocol stages, before announcing her mask $\mu$, Alice could potentially check for Eve's presence. Alice, Bob and Charlie may test for coherence between the two arms by verifying the condition $\{R_H,R_H\}$ (for $H$ photon) or $\{R_V,R_V\}$ (for $V$ photon), that must deterministically give rise to a $D_2$ detection.

\subsection{On the unconditional security of the scheme \label{sec:eveA}}
Now we sketch an unconditional proof of security, where the idea is to consider all choices of Alice and Bob at a quantum level. This leads to a master entanglement in a larger Hilbert space, which makes it easier to analyze the security against Eve. For simplicity, the Bob-injected and Charlie-injected qubits are not taken into account, but this extension can be made in an analogous way. By the method of the larger Hilbert space, Alice's random states are replaced by the quantum superposition
$
\ket{\phi}_A = \frac{\ket{V}_A + \ket{H}_A}{\sqrt{2}},
$
and Bob's \& Charlie's random operations for measurement are given by a coherent superposition of their actions
$
\ket{R_+}_B = \frac{\ket{R_V}_B + \ket{R_H}_B}{\sqrt{2}}$ and
$\ket{R_+}_C = \frac{\ket{R_V}_C + \ket{R_H}_C}{\sqrt{2}}$, respectively.
In the same vein, the initial joint state between Alice, Bob, Charlie and the BS arms is given by
$
\ket{\Psi_0} = \ket{\phi}_A \big(\ket{0}_{D_B} \ket{0}_{D_C} \ket{\phi}_B \ket{\phi}_C\big) \ket{\Psi}_{bc},
$
where $\ket{\Psi}_{bc} = \ket{0}_b\ket{0}_c$ is the initial vacuum state of the BS, whose action is given by
$
\ket{X}_A \rightarrow \frac{\ket{X}_b\ket{0}_c + \ket{0}_b\ket{X}_c}{\sqrt{2}},
$
with $X \in \{H,V\}$ and correspondingly, when light from the arms re-enters the BS, its operation is 
\begin{align}
\ket{X}_b\ket{0}_c &\rightarrow \frac{\ket{D_2^X} +\ket{D_1^X}}{\sqrt{2}}, \nonumber \\
\ket{0}_b\ket{X}_c &\rightarrow \frac{\ket{D_2^X} - \ket{D_1^X}}{\sqrt{2}}.
\end{align}

Then the final state, conditioned on Alice's $D_2$ detections, is 
\begin{widetext}
\begin{align}
\ket{\Psi_1}_{D_2} = \frac{1}{2} \bigg[\ket{D_2^H, R_H, R_H} + \ket{D_2^V, R_V, R_V} + \bigg(\frac{\ket{D_2^H} + \ket{D_2^V}}{4\sqrt{2}}\bigg) \big(\ket{R_H, R_V} + \ket{R_V, R_H} \big)\bigg],
\label{eq:larger}
\end{align}
\end{widetext}
showing the entanglement of the photon between actions by Alice, Bob and Charlie. In the kets, the first, second and third registers represents actions by Alice, Bob and Charlie. Therefore, the state in Eq. (\ref{eq:larger}), restricted to Alice and Bob, reduces to the pure, entangled state:
\begin{align}
\ket{\Psi_2}_{D_2} = \frac{1}{2} \bigg[\ket{D_2^H, R_H} + \ket{D_2^V, R_V} + \sqrt{2}\ket{D_2^+,R_+}\bigg],
\label{eq:monogamy}
\end{align}
where $\ket{D_2^+} \equiv \frac{\ket{D_2^H} + \ket{D_2^V}}{\sqrt{2}}$. The key idea behind invoking the larger Hilbert space is the monogamy of entanglement. This implies that if Alice and Bob can perform Bell state analysis to certify that they possess the entangled state Eq. (\ref{eq:monogamy}) to sufficient degree of certainty, then even without any knowledge of the details if Eve's attack, they can be sure that her state is sufficiently uncorrelated with their private information.

With this, they verify that the statistics of their state verifies Eq. (\ref{eq:monogamy}) and not a mixed state, such as that obtained by tracing out Eve's particle in Eq. (\ref{eq:larger}). It suffices for us to note here that for the above reason, a security analysis on the level of entanglement enables Alice and Bob to obtain unconditional security. Furthermore, composability of the scheme could be an interesting future work. For the remaining section, restricting to a trivial practical scenario, we return to a more conventional method of error analysis. Here, the protocol is considered with the security against certain individual attacks.

\subsection{Security against individual attacks \label{sec:eveB}}

The error induced by Eve can be quantified as follows. Assuming Eve does not collude with involved parties and they remain trusted for the protocol against an information leakage to an eavesdropper, the error here would be the non-$D_1$ detections with respect to Bob's and Charlie's operations, that gives rise to $D_1$ detections. Given that Alice has sent an arbitrary state, we estimate the QBER as
\begin{equation}
e \equiv P(R_VR_V|D_1) + P(R_HR_H|D_1),
\label{eq:error}
\end{equation}
where $$P(R_VR_V|D_1) = \frac{P(D_1|R_VR_V)P(R_VR_V)}{P(D_1)},$$ $$P(R_HR_H|D_1) = \frac{P(D_1|R_HR_H)P(R_HR_H)}{P(D_1)}.$$ 

We know that $P(D_1|R_VR_V)$ and $P(D_1|R_HR_H)$ have two contributions, namely counterfactual error -- $p_1/p_2$ (where a $D_1$ detection happens for a $D_2$, ideally should be zero) and incoherent error -- $\frac{r}{2}$ (error produced due to Bob/Charlie injected bits). Thus let $$ P(D_1|R_VR_V) = p_1 + \frac{r}{2} \;\text{and} \; P(D_1|R_HR_H) = p_2 + \frac{r}{2} .$$

The total number of $D_1$ detections are
\begin{align}
P(D_1) &= P(D_1|R_HR_V) + P(D_1|R_VR_H), \\
&\le \frac{1}{4}(1+r) +p_3, \nonumber
\end{align}
with $p_3$ being the detector error (such as dark count) [Note that $p_3$ is not a factor of $p_1/p_2$, and solely depends on $D_1$ alone]. If Bob's and Charlie's choice of basis for measurement is unbiased, then $P(R_HR_H) = P(R_VR_V) = \frac{1}{4}$. Thus using this in Eq. (\ref{eq:error}), we get
\begin{equation}
e = \frac{p_1+p_2+r}{1+4p_3+r},
\label{eq:erate}
\end{equation}
If $p_1=p_2=p_3=p$, then $$e = \frac{2p+r}{1+4p+r}.$$
This shows the increase of error rate with increase in $p$.

If Eve does IRUD attack, then $p_1$, $p_2$ and $p_3$ gets the value, proportional to the rate of eavesdropping. We find that if Eve's attack rate is $w$ to get polarization information of the Alice sent bits, then the error rate parameters are given by
$p_1 = p_2 = \frac{w}{4}(1+r)$, and
$p_3 = \frac{w}{2}(1+r)$, respectively.
Thus error rate $e$ becomes
\begin{equation}
e = \frac{1}{2}\bigg[\frac{w(1+r) + r}{1 + 2w(1+r) +r}\bigg].
\label{eq:erate}
\end{equation}
Alice's information and Eve's information after eavesdropping are
$I_A = 1-h(e)$ and
$I_E = \frac{w}{2}$, respectively
where $h(\cdot)$ is the Shannon entropy. We know that the parties can have a secure communication if $I_A > I_E$. Thus the secure QDS can happen only when $e_{\rm max} \le 15.3 \%$, when $r=0.01$.

\begin{figure}[ht]
\includegraphics[width=\linewidth]{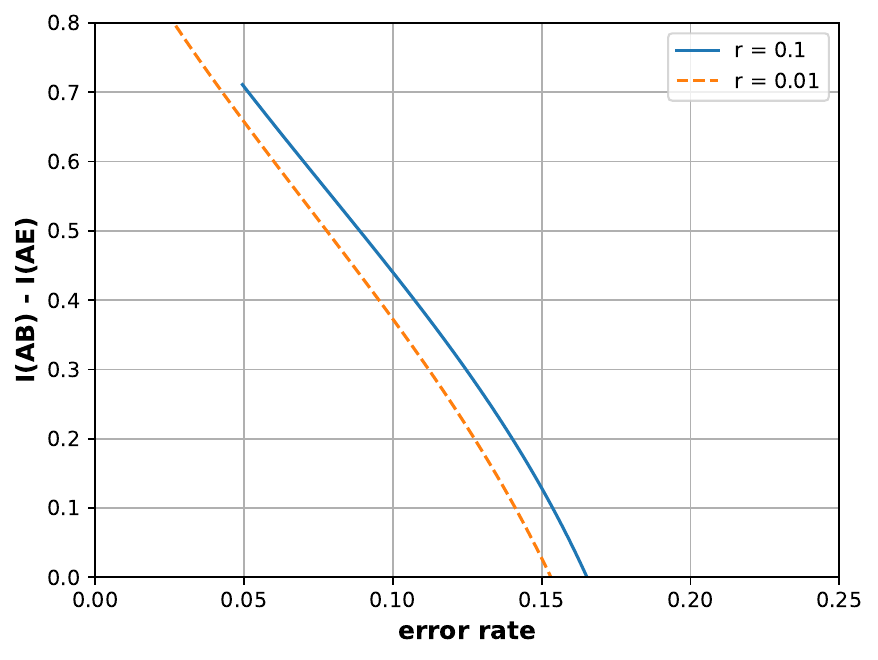}
\caption{Graph of difference in mutual information v/s error rate in the channel, from Eq. (\ref{eq:erate}). The proposed protocol is thus secure only when $e_{\rm max} \le 15.3 \%$ under the considered attack scheme.}
\label{fig:error}
\end{figure}

We have not considered the injected photons by Bob and Charlie in the above security analysis. However, since this is a two-way protocol, Eve may attack the arms twice -- before and after Bob's/Charlie's operation, like W\`ojcik's attack on ping-pong protocol \cite{wojcik2003eavesdropping}.

A similar attack strategy has already been analysed for Noh09 by two of us \cite{rao2021noiseless}, where it is shown that Eve gets complete information of $D_B$ detections, but not of counterfactual ones. Here, if she were to employ such an attack-unattack strategy, she would get information of all the Bob-injected and Charlie-injected bits that lead to $D_1$ detections. A simpler fix to this issue would be to enable Bob and Charlie to flip the reflected polarizations, as in the original work. Then check for coherence between the arms on those instances where they both applied reflect operation and flipped the reflected polarization (a detailed analysis has already been presented in Ref. \cite{rao2021noiseless}). This would restrict Eve to perfectly remove footprint and introduce error. To make the protocol simpler, Alice may throw away all inconsistent polarization detections of both $D_1$ and $D_2$. Thus, the injected bits are present from which security against Alice's repudiation is proven (asymptotically), while also giving security against an eavesdropper. 

\section{Modified protocol \label{sec:mod}}
A practical implementation of the above protocol requires the use of single photons, a somewhat expensive resource when required in sufficiently high rate. Furthermore, it is faced with two further challenges: (a) the generation of redundant bits, i.e., secret, shared bits that the protocol fails to exploit to improve key rate; and (b) a restriction on the secure range owing to the protocol being two-way in nature.

The protocol proposed in Sec. \ref{sec:pro} does not exploit all the secret bits generated between Bob and Charlie. In particular, for a given $D_1$ detection, either one of the parties injected a bit or it was counterfactual in nature. Both could potentially lead to an element of ${\tt Count_{key}}$, and form Alice's signature ${\tt Count_{sig}}$. The table \ref{tab:secretbit} lists the secret key bit and the information shared between the respective parties.

\begin{table}[h]
\centering
\begin{tabular}{|c|c|c|c|}
\hline
Secret bit & Alice & Bob & Charlie \\
\hline
$j$ - polarization & Yes & Yes(i)/No & No/Yes(i) \\
\hline
bit$_{bc}$ - $\{R_j,R_{\overline{j}}\}$ & No & Yes & Yes \\
\hline
\end{tabular}
\caption{Conditioned on a $D_1$ detection, two secret bits are created that are shared between different pairs of parties. Here `Yes(i)' indicates the knowledge of the polarization $j$ due to injection.}
\label{tab:secretbit}
\end{table}

It turns out that one can address all these problems by resorting to an analogous twin-field setup based, one-way QDS protocol that largely retains the logical structure of the above protocol, while eliminating the redundancies. Moreover, a twin-field scheme requires only weak coherent pulses, rather than single-photons. Here we note that this modified protocol still has certain fraction of bits that are wasted, but they are not the secret bits. Conversely, no potential secret bits are thrown away in the sifting.

Now imagine a modified protocol, which, as will be clarified, may be considered as the twin-field, one-way analogue of the above counterfactual, two-way protocol. In the modified scheme, Bob and Charlie prepare and send particles to Alice in the configuration as in Fig. \ref{fig:QDS}. If they are single-photons (as in the primary scheme), then we are led to a two-particle interference. Thus, we employ phase-modulated weak coherent pulses, prepared in $H$ or $V$ polarization. In this case, conditioned on single-photon detections by Alice at detector $D_1$ or $D_2$, we reproduce the same scenario as in the primary scheme. 

Hence if the two incoming pulses are prepared in identical polarization with same phase modulations, then a $D_2$ detection happens. However, $D_1$ detection could happen in the rest of the cases. Interestingly, this can be reduced to the primary scheme, by enabling an announcement by parties if they choose $\pi-$phase. Thus, given no announcement of $\pi-$phase modulation, a detection at detector $D_1$ indicates Alice of different polarization setting, but not the encoding (as in the case of bit$_{bc}$ of Table \ref{tab:secretbit}). This modified protocol potentially eliminates the issue of redundancy and it is one-way in nature. Additionally, the requirement of single-photon states is relaxed as well. Therefore, this protocol could be viewed as the complementary of the primary scheme that is two-way and counterfactual in nature.

But Alice can have the knowledge of bit$_{bc}$ by placing a polarization filter before the beam-splitter. Specifically, Alice could place polarization filters (through which pulse of polarization $j$ passes, and pulse of polarization $\overline{j}$ is blocked) in both the arms. Then a $D_1$ detection, along with a detection at one of the filters, would necessarily reveal the bit$_{bc}$ to Alice. Nevertheless, the security against repudiation due to injection in the primary scheme can be achieved by enabling only one party sending the pulse. Thus, Bob and Charlie can utilize bit$_{bc}$ of these cases, to check against Alice's repudiation. Specifically, Bob and Charlie could exchange information of sent or not sent pulses, thereby performing symmetrization.

By way of making explicit the parallelism and contrast between the primary scheme and the modified protocol, we number the steps of the latter in a way that corresponds sequentially to that of the former scheme. The distribution scheme in the original protocol becomes as follows. 
\begin{enumerate}
\item[(D1)] Alice sets an identical polarization filter for $H$ or $V$ in her end of the communication paths to Bob and Charlie. 
\item[(D2)] Bob and Charlie either send weak coherent signal pulses (signal window) or strong decoy pulses (decoy window), prepared in the basis $H$ or $V$. Each signal pulse is randomly phase-modulated with probability $p_0$ for $0$-phase and $(1-p_0)$ for $\pi$-phase, given $p_0 \gg (1-p_0)$. On fraction $r$ of their detections, they send no pulse towards Alice's station. 
\item[(D3)] The string $\Sigma$ of bits corresponding to Alice's $D_1$ detections and no $\pi-$phase announcement by either of the parties, solely from the signal window, forms the sifted key. Here, Alice's $D_1$ detections are comprised of the cases where Bob and Charlie sending weak pulses of different polarization. The decoy pulses are sent by choosing a specific phase and intensity from a pre-agreed set of values.
\item[(D4)] Bob and Charlie announce part of the data where they sent either the decoy pulses or weak pulses with $\pi-$phase, and together with Alice, they collaboratively estimate the error in the channel using decoy pulses. If it is found to exceed an agreed limit, they abort the protocol.
\item[(D5)] The sifted key $\Sigma = \otimes_1^L \ket{k}\bra{k}$ is obtained from signal pulses and is approximately of length $L \approx N \text{e}^{-|\alpha|^2}\alpha^2$. $\Sigma$ represents Alice's signature ${\tt Count_{sig}}$ and $\{k_1,k_2,\cdots k_L\}$ her private key ${\tt Count_{key}}$. To transmit a one-bit message $m \in \{0,1\}$, Alice publicly announces the corresponding $D_1$ detection coordinates of ${\tt Count_{sig}}$.
\end{enumerate}

The security can be proven from the fact that, conditioned on a single-photon detection at one of the detectors in Alice's station, the states are non-orthogonal, as $\langle \alpha |\beta \rangle = \text{e}^{-|\alpha^2-\beta^2|}$. Specifically, given the lower-bounded single photon detection count $\mathfrak{\underline{n}}_{1}$ and upper-bounded error rate of single-photon states $\mathfrak{\overline{e}}_{1}$, the key length that can be used for Alice's signature is found to be 
\begin{equation}
R \ge \mathfrak{\underline{n}}_{1}[1 - h(\mathfrak{\overline{e}}_{1})],
\label{eq:minent}
\end{equation}
where $k$ corresponds to the input data used by Bob or Charlie to estimate error and $E$ represents the presence of Eve. The two bounds corresponding to signal window in Eq. (\ref{eq:minent}) can be estimated, as below.

The values of $\mathfrak{\underline{n}}_{1}$ and $\mathfrak{\overline{e}}_{1}$ in Eq. (\ref{eq:minent}) correspond to signal states in the scheme. However, we employ standard method of using decoy pulses to estimate them \cite{ma2005practical, jiang2019unconditional}, as follows. This is possible due to the fact that the respective yield (or gain), and error rate of $n$-photon state remains to be same for both signal and decoy pulses \cite{lo2005decoy}.

When decoy mode is chosen, Bob and Charlie choose their polarization setting to be either $j$ (with mean-photon number $|\alpha|^2$) or $\overline{j}$ (with mean-photon number $|\alpha^\prime|^2$ and $\alpha > \alpha^\prime$). If $n_{1}$ is the single-photon detection count of decoy states, we obtain
\begin{align}
n_1 \ge \underline{n}_{1} &= \frac{\chi_{1}}{2\alpha\alpha^\prime (\alpha - \alpha^\prime)} \bigg[ \frac{\text{e}^{-|\alpha^\prime|}|\alpha|^2(n_{0\alpha} + n_{\alpha 0})}{P_{0\alpha}} \nonumber \\
&-\frac{\text{e}^{-|\alpha|}|\alpha^\prime|^2(n_{0\alpha^\prime} + n_{\alpha^\prime 0})}{P_{0\alpha^\prime}} 
- \frac{2({|\alpha|^2} - {|\alpha^\prime|^2})n_{00}}{P_{00}} \bigg], 
\end{align}
where the first two terms in RHS correspond to the detection at $D_1$ for either polarization setting $j$ or $\overline{j}$ by Alice, respectively, the third term for the setting that blocks both the pulses (detections from dark counts alone) and $\chi_{1} = \sum_{y=\alpha ,\alpha^\prime} (yP_{0y} \text{e}^{-y})$ is the probability of single-photon events, with $P_{0y}$ indicating the probability with which Bob sent pulse was blocked and Charlie sent pulse $y$ was detected at Alice's station. Similarly, if $e_1$ indicates the error rate of single-photon states of the decoy states,
\begin{align}
e_1 \le \overline{e}_1 = \frac{\chi_{1}}{n_1 (\alpha - \alpha^\prime)}\big[e_1^{(1)} + e_1^{(2)}\big].
\label{eq:derror}
\end{align}

Here the first term in RHS corresponds to the case wherein a $D_1$ detection happens when the identical pulses are sent and the second term indicates the detections at both- the filter and one of the detectors. This can be estimated using
\begin{align}
e_1^{(1)} &= \frac{\text{e}^{-|\alpha|}|\alpha|^2(n_{\alpha(00)} + n_{\alpha(\pi \pi)})}{P_{\alpha(xx)}} \nonumber \\
&- \frac{\text{e}^{-|\alpha^\prime|}|\alpha^\prime|^2(n_{\alpha^\prime (00)} + n_{\alpha^\prime (\pi \pi)})}{P_{\alpha^\prime (xx)}}, \nonumber \\
e_1^{(2)} &= \frac{\text{e}^{-|j|}j^2(n_{fj} + n_{jf})}{P_{fj}}.
\end{align}

Here $e_1^{(2)}$ is the total number of detections, summed over both the polarizations $j \in \{\alpha,\alpha^\prime\}$, $P_{y(xx)}$ denotes both sending pulse $y$ with phase $x$, $n_{fj}$ \& $n_{jf}$ indicate the detection at a filter and a detector and $n_{ab}:= n\pm \sqrt{\frac{n\ln (\epsilon_F^{-1})}{2}}$ is the observed value due to statistical fluctuations by Hoeffding inequality \cite{hoeffding1994probability}, with $\epsilon_F$ denoting the failure probability. Note that these estimations are for decoy states, and to estimate $\mathfrak{\underline{n}}_{1}$ and $\mathfrak{\overline{e}}_{1}$ (of signal states) we use Serfling inequality as, 
\begin{align}
\mathfrak{n}_{1} \ge \mathfrak{\underline{n}}_{1} &= n_1\frac{L}{2\mathfrak{n}_z} - \Gamma(\mathfrak{n}_z, \frac{L}{2}, \epsilon_F), \nonumber \\
\mathfrak{e}_{1} \le \mathfrak{\overline{e}}_{1} &= e_1\frac{1}{n_1} + \Gamma(\mathfrak{\underline{n}}_{1}, n_1, \epsilon_F), 
\label{eq:serf}
\end{align}
where $\Gamma(a,b,c) = \sqrt{(a-b+1)b \ln(c^{-1})/(2a)}$ is the factor of sampling without replacement in Serfling's inequality.

\begin{figure}[ht]
\includegraphics[width=\linewidth]{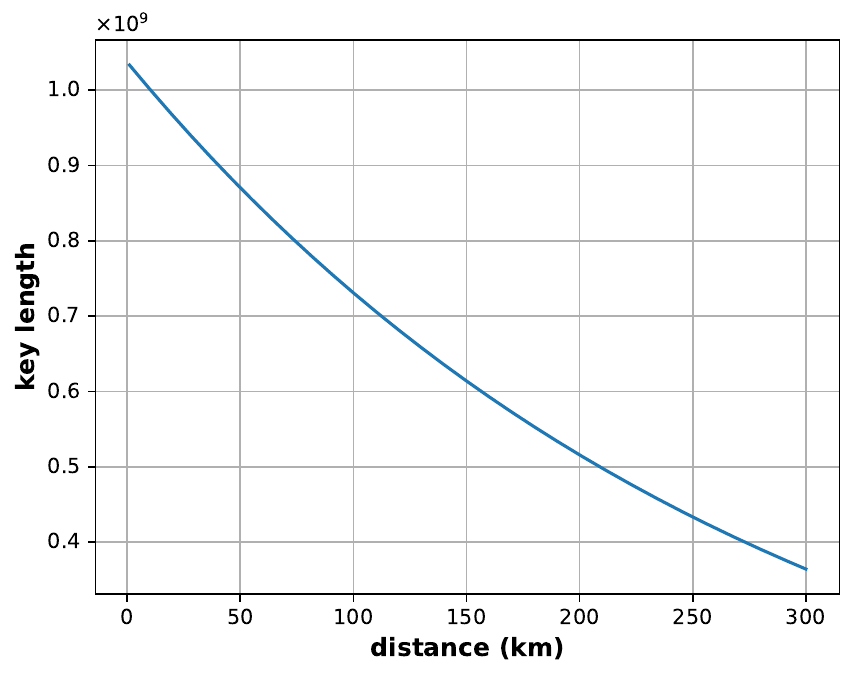}
\caption{Graph of signature length v/s distance. Here the values for numerical analysis are as follows: $N = 10^{10}$, misalignment error $e_m = 0.03$, dark count detection probability $p_d = 10^{-7}$, statistical fluctuations $\epsilon_{F} = 10^{-12}$, transmittivity $\eta = 0.5 \times 10^{-\frac{\lambda}{200}}$ (with $\lambda = 0.3$ dB/km) and the two amplitudes of $H$ and $V$ polarization pulses are $\alpha = 0.15$ and $\alpha^\prime = 0.1$, respectively.}
\label{fig:keylength}
\end{figure}

In addition to the error rate of single-photon states $\mathfrak{\overline{e}}_{1}$ estimated using the decoy states as in Eq. (\ref{eq:serf}), one can also estimate the overall quantum bit error rate $\mathfrak{E}_{tot}$, as 
\begin{equation*}
\mathfrak{E}_{tot} = \frac{\mathfrak{n}_{ss}+\mathfrak{n}_{00}}{\mathfrak{n}_{tot}},
\end{equation*}
where $\mathfrak{n}_{ss}$ (resp., $\mathfrak{n}_{00}$) corresponds to the number of events of Bob and Charlie both sending pulses (resp., neither sending a pulse) in signal window, and Alice announcing a $D_1$ detection, and $\mathfrak{n}_{tot}$ being number of successful detections in signal window. This can effectively be used for classical post processing of error correction in Eq. (\ref{eq:minent}) as,
\begin{equation}
R \ge \mathfrak{\underline{n}}_{1}[1 - h(\mathfrak{\overline{e}}_{1})] - \mathfrak{n}_{tot}fh(\mathfrak{E}_{tot}),
\end{equation}
with $f$ representing the error correction efficiency factor. The various quantities of above equations are suitably estimated as in Refs. \cite{jiang2019unconditional, zhang2021twin} and the protocol is numerically simulated, as given in Fig. \ref{fig:keylength}.

\section{Discussion \& Conclusion \label{sec:conc}}
We have presented a quantum digital signature scheme whose security is guaranteed by quantum counterfactuality. The proposed scheme is different from other QDS protocols in that we make use of only orthogonal states and require neither the quantum memory nor the multiport \cite{gottesman2001quantum, andersson2006experimentally}.

In the proposed QDS scheme, Alice sends the quantum states to Bob and Charlie, along with the detection coordinates corresponding to her signature ${\tt Count_{sig}}$, in the distribution stage. The sending of the message, with her private key ${\tt Count_{key}}$, happens only in the messaging stage. However, the security against Alice's repudiation comes from state comparison as well as injected bits. This is different from the other QDS schemes, the security against repudiation is either through state comparison by Bob and Charlie \cite{gottesman2001quantum} or by their symmetrization of states \cite{clarke2012experimental}. It is easily implementable with the presently available technology. The bounds presented give evidence for security concerns against repudiation and signature forging. We have also considered an eavesdropper's attack and security regarding that has been addressed in detail. We point out that the injection of photons is equivalent to the principle of symmetrization employed in various other QDS protocols. Hence, the protocol can be thought of belonging to the same class of such QDS protocols. In particular, the asymptotic security of the present scheme is in line with conventional security addressed in QDS schemes. However, an unconditional security is yet to be established for the present scheme.

Incidentally, though a beam-splitter of a Michelson interferometer is present in the protocol, and thereby involves a multiport for the scheme, the important distinction from other multiport-based schemes is that we need only one BS and it is associated with the 
sender alone. The QKD component itself involves a BS, and no further extension is required for the QDS scheme. Hence the assertion that the protocol requires neither the quantum memory nor the multiport to prove the security.

We note here that the essentiality of the error correction in the primary scheme is not discussed in various other QDS protocols \cite{dunjko2014quantum, wallden2015quantum, amiri2016secure}. Specifically, the impossibility of Bob using information due to symmetrization, to forge Alice's signature during forwarding, is paramount in proving the security. Therefore, we employ the error correction in our scheme. However, another slight modification, at the cost of greater quantum resources, could potentially detect Bob's forgery.

In particular, we propose a variant of the primary QDS scheme, wherein Alice prepares the $N$-qubit composite state $\ket{\varphi_i}$ for $\gamma >1$ rounds and $i \in \{\gamma\}$. The corresponding set of $D_1$ detections of $i$-th round be $\Sigma_i$, given Bob and Charlie could change their operations at each round. After $\gamma$ rounds, Alice chooses her signature $\Sigma$ and announces the corresponding $D_1$ detections. Therefore, at the end of distribution stage, Bob's and Charlie's knowledge of $\ket{\varphi}$ would be close to each other. 

Suppose the $j$-th bit ($j \in \{1,2,\cdots,N\}$) resulted in a $D_1$ detection due to Bob's injection in the $i$-th round. He would flip the bit when forwarding the message to Charlie. However, Charlie too could know the polarization of the $j$-th bit from $(k \ne i)$-th round. Thus, the probability for Bob to successfully forge decreases as $\gamma$ increases. Thus, the error correction is not required in this modified protocol. Additionally, the injection too could in principle be relinquished, as the security against repudiation might be proven from an identical argument as above.

It is important to note that Eve could get information from an optimal quantum cloning machine, as the state $\ket{\varphi}$ is identical in each round \cite{gisin1997optimal, bruss1998optimal2}. The guessing probability of state discrimination could be potentially improved \cite{zhang1999general, barnett2009quantum}, estimated using Helstrom bound \cite{helstrom1969quantum}. In various other QDS schemes too, multiple copies of identical states are sent by the sender to two or more parties. This issue of multiple copies being used could further impact on multiparty QDS schemes \cite{xu2014novel, arrazola2016multiparty} as well. However, in the context of present work, this can be thwarted by varying the composite state $\ket{\varphi_i}$ in each round, such that the fidelity between the two composite states (of any two rounds) is close (but not equal) to unity.

The phenomenon of counterfactual security \cite{rao2021noiseless} in QKD is inconsequential here, as only the $D_1$ detections are used for generation of Alice's signature. If one uses all of $D_B$ and $D_C$ detections, then the QDS protocol becomes insecure as one party (forwarder) can cheat by flipping those bits. Also from the noiseless attack, Eve can potentially get full information of $D_B$ and $D_C$ detections. But if she eavesdrops for counterfactual detections, she must produce error as she selectively cannot attack on those instances.

We have also proposed a modified QDS protocol based on the idea of twin-field cryptography. Specifically, another set of secret bits shared between Bob and Charlie in the primary protocol were unused in the primary protocol, and therein lies the possibility of utilizing that using TF-QKD setup. The primary advantage of this would be the nature of protocol being one-way, and using coherent states for the key generation. We note that the framework of the protocol remains to be same. It is worth pointing out here that the idea used above, of converting a two-way scheme to its one-way equivalent by replacing the single-photon qubits in the original scheme by weak coherent pulses in the latter, can also be applied in a QKD situation, e.g., to the Noh09 protocol \cite{noh2009counterfactual} for counterfactual QKD.

The proposed scheme can be potentially generalized to a multiparty scenario with multiple forwarders. The quantum part of the protocol would be similar, with Alice sending states through an $n$-input $n$-output beam-splitter. The potential key length of Alice's signature would increase for such a setup, as the probability for a counterfactual detection increases. However, the underlying security would be from the injected bits, as in the proposed scheme. The limitation of these protocols is that, similar to other QDS protocols, this is one-time secure protocol. In other words, for every new message, the parties must perform a run of the protocol. Eavesdropper can launch more powerful incoherent attacks, but the nature of security remains same - she cannot remove her footprint if she extracts information.

Finally, note that in any QDS protocol, the assumption that the receiver or verifier is trusted follows the tradition in the classical DS literature. It is an interesting question what modifications are to be made to a given (counterfactual) QDS protocol to ensure security against Charlie's dishonesty. We leave this as an open question.

\acknowledgments

The authors thank Sandeep Mishra and Kishore Thapliyal for the useful discussions. VNR, AP and RS acknowledge partial financial support from the Interdisciplinary Cyber Physical Systems (ICPS) program of DST, India, Grant No. DST/ICPS/QuST/Theme-1/2019/14 (Q80). SU thanks IIT Madras for the support through the Institute Post-Doctoral Fellowship. VNR also acknowledges the support and encouragement from the Admar Mutt Education Foundation and Manipal Academy of Higher Education during the time when this work was initiated.

\bibliography{references}

\appendix

\section{Decoy state}

Consider the case wherein Alice sends quantum states with probability distribution $P_i$, with $i$ being the photon number. Then the total gain is given by,
\begin{align}
Q = \sum_{i=0}^{\infty} Y_i P_i,
\end{align}
where $Y_i$'s are the yield (detection counts) of respective photon number states. Here, the yield $Y_i$ for a given photon number state remains identical for any quantum state (for a given setup), as they depend only on the transmittivity and detection efficiency of the detectors. Therefore, this property is employed for QKD, where weak coherent states are sent in place of single-photon states, and consequently, coherent states of other amplitudes can act as decoy states. Since the yields are independent of signal/decoy states, we estimate the gain of signal pulses using decoy states, as in the former of Eq. (\ref{eq:serf}).

Similarly, if the total error in the channel can be estimated as,
\begin{align}
E_{tot}Q = \sum_{i=0}^{\infty} e_i Y_i P_i,
\end{align}
where $E_{tot}$ is the weighted average of the QBERs $e_i$, corresponding to the photon number states $i$. Additionally, $e_1$ is the QBER corresponding to single-photon states, as in the Eq. (\ref{eq:derror}). Hence the error rate of single-photon states $e_1$ is estimated using decoys, which in turn is used to find the same of signal pulses $\mathfrak{e}_{1}$, as in the latter of Eq. (\ref{eq:serf}). Therefore, comparing the BB84 protocol and the present case, $E_{tot}$ would be estimated using test bits of $Z$-basis (a part of sifted bits) in the former.

\end{document}